\documentclass[sigconf]{acmart}
\usepackage{booktabs}
\usepackage{multirow}
\usepackage{subcaption}
\colorlet{bestmodel}{blue!5}

\newcommand{\da}{$^\dagger$}
\newcommand{\dda}{$^\ddagger$}

\setcopyright{acmcopyright}
\copyrightyear{2020}
\acmYear{2020}
\acmDOI{10.1145/1122445.1122456}

\acmConference[JCDL '20]{JCDL '20: ACM/IEEE Joint Conference on Digital Libraries}{August 01--05, 2020}{Xi'an, Shaanxi, China}

\acmBooktitle{JCDL '20: ACM/IEEE Joint Conference on Digital Libraries, August 01--05, 2020, Xi'an, Shaanxi, China}
\acmISBN{978-1-4503-XXXX-X/20/06}

\begin{document}

\title{Large-Scale Evaluation of Keyphrase Extraction Models}

\author{Ygor Gallina}
\email{ygor.gallina@univ-nantes.fr}
\affiliation{%
  \institution{LS2N, Université de Nantes}
  \city{Nantes}
  \country{France}
}

\author{Florian Boudin}
\email{florian.boudin@univ-nantes.fr}
\affiliation{%
  \institution{LS2N, Université de Nantes}
  \city{Nantes}
  \country{France}
}

\author{Béatrice Daille}
\email{beatrice.daille@univ-nantes.fr}
\affiliation{%
  \institution{LS2N, Université de Nantes}
  \city{Nantes}
  \country{France}
}


\begin{abstract}
Keyphrase extraction models are usually evaluated under different, not directly comparable, experimental setups.
As a result, it remains unclear how well proposed models actually perform, and how they compare to each other.
In this work, we address this issue by presenting a systematic large-scale analysis of state-of-the-art keyphrase extraction models involving multiple benchmark datasets from various sources and domains.
Our main results reveal that state-of-the-art models are in fact still challenged by simple baselines on some datasets.
We also present new insights about the impact of using author- or reader-assigned keyphrases as a proxy for gold standard, and give recommendations for strong baselines and reliable benchmark datasets.
\end{abstract}



\maketitle

\section{Introduction}

Keyphrases are single or multi-word lexical units that represent the main concepts in a document~\cite{evans-zhai:1996:ACL}.
They are particularly useful for indexing, searching and browsing digital libraries~\cite{barker1972comparative,zhai-1997-fast,Gutwin:1999:IBD:338985.338996,witten2009build}, and have proven themselves as effective features in many downstream natural language processing tasks~\cite{hulth-megyesi:2006:COLACL,litvak-last:2008:MMIES,berend:2011:IJCNLP-2011}.
Still, most documents do not have assigned keyphrases, and manual annotation is simply not a feasible option~\cite{Mao2017}.
There is therefore a great need for automated methods to assign relevant keyphrases to documents.

Automatic keyphrase extraction\footnote{Also referred to as keyphrase generation or keyphrase annotation.} -- that is, the task of extracting keyphrases either from the content of the document or from a controlled vocabulary -- has received much attention from the research community~\cite{kim-EtAl:2010:SemEval,Keyphrase:2015,augenstein-EtAl:2017:SemEval}.
Thus, many keyphrase extraction models were proposed over the last years, ranging from early statistics-based models~\cite{Witten:1999:KPA:313238.313437}, to popular graph-based ranking models~\cite{mihalcea-tarau:2004:EMNLP}, and recent neural models~\cite{P17-1054}.
However, because of the great discrepancies in experimental setups among past studies, it is very difficult to compare and contrast the effectiveness of these models, and even more so to assess the progress of the field as a whole.

More specifically, we observe striking differences in how models are parameterized, evaluated and compared in previous work.
To name just a few examples, experiments are most often conducted on different benchmark datasets, all of which differ in domain, size, language or quality of the gold standard (that is, reference keyphrases supplied by authors, readers or professional indexers).
This not only makes the reported results hard to contrast, but also has a profound impact on trained model performance~\cite{gallina-etal-2019-kptimes}.
%
%
In addition, and since there is no consensus as to which evaluation metric is most reliable for keyphrase extraction~\cite{zesch-gurevych:2009:RANLP09,reading32266,hasan-ng:2014:P14-1}, diverse measures are commonly seen in the literature, thus preventing any further direct comparisons.
Moreover, the evaluation of missing keyphrases -- that is, gold keyphrases that do not occur in the content of the document -- is still an open question and there is little agreement on whether they should be included or not~\cite{kim-EtAl:2010:SemEval}.

We strongly believe that this lack of empirical rigor is a real hindrance to progress on keyphrase extraction, and that a systematic comparison of existing models under the same conditions is needed to fully understand how they actually perform.
In this work, we resolve this issue by conducting the first large-scale study on automatic keyphrase extraction.
More precisely, we present an extensive comparative analysis of state-of-the-art keyphrase extraction models involving 9 benchmark datasets from various domains.
To ensure controlled, fair and reliable experiments, we embarked upon the difficult process of re-implementing all of the models presented in this paper\footnote{Link to the code will appear here after the review period.} and pre-processing the datasets in a unified and systematic way\footnote{Link to the datasets will appear here after the review period.}.

%
%

Using these new large-scale experimental results, we seek to better understand how well state-of-the-art models perform across sources, domains and languages. 
We also go further than prior work and investigate the following research questions:
\begin{enumerate}
    \item How much progress have we made on keyphrase extraction since early models?
    \item What is the impact of using non-expert gold standards, that is, author- or reader-assigned keyphrases, when training and evaluating keyphrase extraction models?
    \item Which baselines and benchmark datasets should be included in future work for a better understanding of the pros and cons of a newly proposed model?
\end{enumerate}

\section{Benchmark Datasets}
\label{sec:datasets}

Benchmark datasets for evaluating automatic keyphrase extraction cover a wide range of sources ranging from scientific articles and web pages to twitter and email messages.
We collected 9 of the most widely used datasets which we believe are representative of the different sources and domains found in previous work.
Detailed statistics for each selected dataset are shown in Table~\ref{tab:datasets}.
They are grouped into three categories that are outlined below:

\begin{description}

\item[Scientific articles]
Among the selected datasets, three are composed of full-text scientific publications: ACM~\cite{krapivin2009large} and SemEval~\cite{kim-EtAl:2010:SemEval} about computer science, and PubMed~\cite{schutz2008keyphrase} from the medical domain.
Not surprisingly, they contain only a small number of documents due to copyright reasons.
These datasets provide author-assigned keyphrases which serve as a reasonable, but far from perfect, proxy for expert annotations.
%
In the case of SemEval, student annotators were hired to extend gold annotation labels.
    
\item[Paper abstracts]
Scientific abstracts, often referred to as bibliographic records, are arguably the most prevalent documents for benchmarking keyphrase extraction.
They are readily available in great quantities and come with author-assigned keyphrases that can be used as gold standard.
We gathered three datasets, all dealing with the computer science domain: Inspec~\cite{Hulth:2003:EMNLP}, WWW~\cite{caragea-EtAl:2014:EMNLP2014} and KP20k~\cite{P17-1054}.
It is worth noting that with more than half a million documents, KP20k is the largest dataset to date and one of the few that is large enough to train neural models.

\item[News articles] 
News texts are the last source of documents present among the collected datasets.
Similar to paper abstracts, online news are available in large quantities and can be easily mined from the internet.
%
%
We selected the following three datasets: DUC-2001~\cite{Wan:2008:SDK:1620163.1620205}, 500N-KPCrowd~\cite{MARUJO12.672} and KPTimes~\cite{gallina-etal-2019-kptimes}.
The first two datasets provide reader-assigned keyphrases, while KPTimes supplies indexer-assigned key-phrases extracted from metadata and initially intended for search engines.
It is interesting to observe that only two datasets in our study, namely Inspec and KPTimes, provide gold keyphrases annotated by professional indexers.
    
\end{description}

\begin{table}[ht!]
\centering
\setlength{\tabcolsep}{4.5pt}
    \begin{tabular}{@{}rcrrrrr}
    
    \cmidrule[1pt]{1-7}
        \textbf{Dataset} &
        \textbf{Ann.} &
        \textbf{Train} &
        \textbf{Test} &
        \textbf{\#words} &
        \textbf{\#kp} &
        \textbf{\%abs} \\
    \cmidrule[.5pt]{1-7}

    PubMed~\cite{schutz2008keyphrase}         & $A$        & -   & 1\,320 & 5\,323  & 5.4  & 16.9 \\
    ACM~\cite{krapivin2009large}              & $A$        & -   & 2\,304 & 9\,198  & 5.3  & 16.3 \\
    SemEval~\cite{kim-EtAl:2010:SemEval} & $A \cup R$ & 144 & 100    & 7\,961  & 14.7 & 19.7 \\
    
    \cmidrule{5-7} 
    
    \multicolumn{4}{r}{\textit{Scientific articles (avg.)}} &  7~494 &  8.5 & 17.6 \\
    
    \cmidrule[.5pt]{1-7}
    
    Inspec~\cite{Hulth:2003:EMNLP}         & $I$ & 1\,000 & 500    & 135 & 9.8 & 22.4 \\
    WWW~\cite{caragea-EtAl:2014:EMNLP2014} & $A$ & -      & 1\,330 & 164 & 4.8 & 52.0 \\
    KP20k~\cite{P17-1054}                  & $A$ & 530K & 20K & 176 & 5.3 & 42.6 \\
    
    \cmidrule{5-7} 
    \multicolumn{4}{r}{\textit{Paper abstracts (avg.)}} &  158 &  6.6 & 39.0 \\
    
    \cmidrule[.5pt]{1-7}
    
    DUC-2001~\cite{Wan:2008:SDK:1620163.1620205} & $R$ & -   & 308 & 847 & 8.1    &  3.7 \\
    KPCrowd~\cite{MARUJO12.672}                  & $R$ & 450 & 50  & 465 & 46.2   & 11.2 \\
    KPTimes~\cite{gallina-etal-2019-kptimes}     & $I$ & 260K & 10K & 921 & 5.0 &  54.7 \\
    
    \cmidrule{5-7} 
    \multicolumn{4}{r}{\textit{News articles (avg.)}} &    744 &  19.8 & 23.2 \\

    \cmidrule[1pt]{1-7}
    
    \end{tabular}
\caption{Statistics of the datasets. Gold annotation is supplied by authors ($A$), readers ($R$) or professional indexers ($I$). The number of documents in the training and testing splits are shown. The average number of keyphrases (\#kp) and words (\#words) per document, and the ratio of missing keyphrases (\%abs) are computed on the test set.} 
\label{tab:datasets}
\end{table}

Datasets containing scientific articles or abstracts rely primarily on author-assigned keyphrases as gold standard.
They therefore exhibit similar properties for the average number of ground truth keyphrases per document ($\approx5$).
On the other hand, articles are on average significantly longer than abstracts ($\approx 7500$ words vs. $\approx 160$ words respectively) and consequently reveal a much smaller fraction of missing keyphrases ($\approx 18\%$ vs. $\approx 39\%$ respectively). 
Datasets with reader-assigned keyphrases exhibit the lowest numbers of missing keyphrases, which can be explained by the fact that readers appear to produce gold-standard annotations in an extractive fashion~\cite{10.1007/978-3-319-19548-3_21}.
We also confirmed this empirically by computing the ratio of missing keyphrases in the author-assigned ($24\%$) and reader-assigned ($17.5\%$) gold annotations of the SemEval dataset.

In contrast, the opposite trend is observed for KPTimes that comes with gold standards annotated by professional indexers and that shows the highest percentage of missing keyphrases ($54.7\%$).
This indicates the the more abstractive nature of indexer-assigned keyphrases.
Put differently, it is known that non-expert annotations are less constrained and may include seldom-used variants or misspellings~\cite{DBLP:conf/icwsm/SoodOHB07}, whereas indexers strive to rely on a consistent terminology and assign the same keyphrase to all documents for a given topic, even when it does not occur in these documents.

To investigate this further, we looked at how many variants of an index term, in this case ``\textit{artificial neural network}'', could be found in the author-assigned keyphrases of KP20k.
All in all, we found dozens of variants for this term, including ``\textit{neural network}'', ``\textit{neural network (nns)}'', ``\textit{neural net}'', ``\textit{artificial neural net}'' or ``\textit{nn}''.
This apparent lack of annotation consistency intuitively has two consequences: 1) it makes it harder for supervised approaches to learn a good model, 2) it makes automatic evaluation much less reliable as it is based on exact string matching.

It is important to stress that datasets containing scientific articles may contain noisy texts. 
Indeed, most articles were automatically converted from PDF format to plain text and thus are likely to contain irrelevant pieces of text (e.g.~muddled sentences, equations).
Previous work show that noisy inputs undermine the overall performance of keyphrase extraction models~\cite{boudin-mougard-cram:2016:WNUT}.
In this study, we do not insist on a perfect input and we are aware that reported results may be improved with an increase in pre-processing effort. 

\section{Models}

Roughly speaking, previous works on keyphrase extraction can be divided into two groups depending on whether they adopt a supervised learning procedure or not.
%
%
This section starts by introducing the baselines we will use in our experiments, and then proceeds to describe the state-of-the-art keyphrase extraction models we re-implemented sorted into the aforementioned two groups.


\subsection{Baselines}

Having strong baselines to compare with is a prerequisite for contrasting the results of proposed models.
In previous studies, various baselines were considered, complicating the analysis and interpretation of the reported results.
Our stance here is to establish three baselines, each associated with a particular feature that is commonly used in keyphrase extraction models.
All baselines are also unsupervised, allowing their use and performance analysis on any of the benchmark datasets

Keyphrase position is a strong signal for both unsupervised and supervised models, simply because texts are usually written so that the most important ideas go first~\cite{marcu:1997:ACL}.
In single document summarization for example, the lead baseline --that is, the first sentences from the document--, while incredibly simple, is still a competitive baseline~\cite{kedzie-mckeown-daumeiii:2018:EMNLP}.
Similar to the lead baseline, we propose the \textbf{FirstPhrases} baseline that extracts the first $N$ keyphrase candidates from a document.
We are not aware of any previous work reporting that baseline, yet, as we will see in \S\ref{sec:results}, it achieves remarkably good results.

Graph-based ranking models for keyphrase extraction are, perhaps, the most popular models in the literature.
Therefore, as a second baseline, we use \textbf{TextRank}~\cite{mihalcea-tarau:2004:EMNLP}, which weights keyphrase candidates using a random walk over a word-graph representation of the document.
In a nutshell, TextRank defines the importance of a word in terms of how it relates to other words in the document, and ranks candidates according to the words they contain.

%
%
%

The third baseline, \textbf{TF$\times$IDF}~\cite{SALTON1988513}, have been repeatedly used in previous comparative studies~\cite[\textit{inter alia}]{kim-EtAl:2010:SemEval,P17-1054}.
In contrast with the other two baselines that do no require any resources whatsoever (beyond the document itself), TF$\times$IDF makes use of the statistics collected from unlabelled data to weight keyphrase candidates.
As such, it often gives better results, in some cases even on par with state-of-the-art models~\cite{ye-wang:2018:EMNLP}. 

\subsection{Unsupervised models}

Annotated data are not always available or easy to obtain, which motivates the further development of unsupervised models for keyphrase extraction.
Besides, looking back at previous work, most attempts to address this problem employ unsupervised approaches.
%
In this study, we selected three recent state-of-the-art models based on their reported performance. 

The first model we investigate is \textbf{PositionRank}~\cite{florescu-caragea:2017:Long}, a graph-based model that incorporates two features (position and frequency) into a biased PageRank algorithm.
This model operates at the word level, and assigns a score to each candidate using the sum of its individual word scores.
As such, it suffers from over-generation errors\footnote{These errors occur when a model correctly outputs a keyphrase because it contains an important word, but at the same time erroneously predicts other keyphrases because they contain the same word.}~\cite{hasan-ng:2014:P14-1}, but still achieves good performance on short texts.

The second model we consider, \textbf{MPRank}~\cite{boudin:2018:N18-2}, relies on a multipartite graph representation to enforce topical diversity while ranking keyphrase candidates.
It includes a mechanism to incorporate keyphrase selection preferences in order to introduce a bias towards candidates occurring first in the document.
%
%
MultipartiteRank was shown to consistently outperform other unsupervised graph-based ranking models. 


Both aforementioned models only exploit the document itself to extract keyphrases.
The third model we include, \textbf{EmbedRank}~\cite{bennanismires-EtAl:2018:K18-1}, leverages sentence embeddings for ranking keyphrase candidates.
%
%
Candidates are weighted according to their cosine distance to the document embedding, while diversity in the selected keyphrases is promoted using Maximal Marginal Relevance (MMR)~\cite{goldstein-carbonell:1998:TIPSTER98}.
Despite its simplicity, this model was shown to outperform other unsupervised models on short texts (abstracts and news).

\subsection{Supervised models}

Supervised models can be further divided into two categories, depending on whether they rely on a neural network or not.


Traditional supervised models treat the keyphrase extraction problem as a binary classification task.
Here, we include such a model, namely \textbf{Kea}~\cite{Witten:1999:KPA:313238.313437}, in order to precisely quantify the performance gap with recent neural-based models. 
KEA uses a Naive Bayes classifier trained on a set of only two handcrafted features we have elected as baseline features: the TF$\times$IDF score of the candidate and the normalized position of its first occurrence in the document.
%
%
Previous work has reported confusing and conflicting results\footnote{On SemEval, \cite{P17-1054} report an F@10 score of $2.6$ while \cite{boudin:2016:COLINGDEMO} report a score of $19.3$.} for Kea, raising questions about how it actually performs. 

%
%

Neural models for keyphrase extraction rely on an encoder-decoder architecture~\cite{cho-EtAl:2014:EMNLP2014,NIPS2014_5346} with an attention mechanism~\cite{bahdanau2014neural,luong-pham-manning:2015:EMNLP}.
Training these models require large amounts of annotated training data, and is therefore only possible on the KP20k and KPTimes datasets.
The second supervised model we include in this study is \textbf{CopyRNN}~\cite{P17-1054}, an encoder-decoder model that incorporates a copying mechanism~\cite{gu-EtAl:2016:P16-1} in order to be able to predict phrases that rarely occur.
When properly trained, this model was shown to be very effective in extracting keyphrases from scientific abstracts.


The third supervised model we use, \textbf{CorrRNN}~\cite{chen-EtAl:2018:EMNLP9}, extends the aforementioned model by introducing correlation constraints.
It employs a coverage mechanism~\cite{tu-etal-2016-modeling} that diversifies attention distributions to increase topic coverage, and a review mechanism to avoid generating duplicates.
As such, it produces more diverse and less redundant keyphrases.

Note that only neural models have the ability to generate missing keyphrases, which in theory gives them a clear advantage over the other models.

\section{Experimental settings}
\label{sec:expe}

In addition to the variation in the choice of benchmark datasets and baselines, there are also major discrepancies in parameter settings and evaluation metrics between previous studies.
For example, there is no point in contrasting the results in~\cite{P17-1054}, \cite{florescu-caragea:2017:Long} and \cite{teneva-cheng:2017:Short}, three papers about keyphrase extraction published in the same year at ACL, since neither benchmark datasets, parameter settings nor evaluation metrics are comparable.
To address this problem, we use the same pre-processing tools, parameter settings and evaluation procedure across all our experiments.

\begin{table*}[ht!]
    \centering
    \resizebox{\textwidth}{!}{
    \begin{tabular}{r c@{\hspace*{2mm}}c c@{\hspace*{2mm}}c c@{\hspace*{2mm}}c | c@{\hspace*{2mm}}c c@{\hspace*{2mm}}c c@{\hspace*{2mm}}c | c@{\hspace*{2mm}}c c@{\hspace*{2mm}}c c@{\hspace*{2mm}}c}
    
        ~ &
        \multicolumn{6}{c}{\textit{Scientific articles}} &
        \multicolumn{6}{c}{\textit{Paper abstracts}} &
        \multicolumn{6}{c}{\textit{News articles}}
        \\
        
        \cmidrule(lr){2-7} \cmidrule(lr){8-13} \cmidrule(lr){14-19}
    
        ~ &
        \multicolumn{2}{c}{\textbf{PubMed}} &
        \multicolumn{2}{c}{\textbf{ACM}} &
        \multicolumn{2}{c}{\textbf{SemEval}} &
        \multicolumn{2}{c}{\textbf{Inspec}} &
        \multicolumn{2}{c}{\textbf{WWW}} &
        \multicolumn{2}{c}{\textbf{KP20k}} &
        \multicolumn{2}{c}{\textbf{DUC-2001}} &
        \multicolumn{2}{c}{\textbf{KPCrowd}} &
        \multicolumn{2}{c}{\textbf{KPTimes}} \\

        \cmidrule(lr){2-3} \cmidrule(lr){4-5} \cmidrule(lr){6-7}
        \cmidrule(lr){8-9} \cmidrule(lr){10-11} \cmidrule(lr){12-13}
        \cmidrule(lr){14-15} \cmidrule(lr){16-17} \cmidrule(lr){18-19}
        
        \\[-1.5em]
        
        \textbf{Model} &
        \small{$\text{F}@10$} & \small{MAP} & \small{$\text{F}@10$} & \small{MAP} & \small{$\text{F}@10$} & \small{MAP} &
        \small{$\text{F}@10$} & \small{MAP} & \small{$\text{F}@10$} & \small{MAP} & \small{$\text{F}@10$} & \small{MAP} &
        \small{$\text{F}@10$} & \small{MAP} & \small{$\text{F}@10$} & \small{MAP} & \small{$\text{F}@10$} & \small{MAP} \\
        
        \midrule

		FirstPhrases &
		15.4 & 14.7 & 13.6 & 13.5 & 13.8 & 10.5 &
		29.3 & 27.9 & 10.2 & \phantom{0}9.8 & 13.5 & 12.6 &
		24.6 & 22.3 & 17.1 & 16.5 & \phantom{0}9.2 & \phantom{0}8.4 \\

		TextRank &
		\phantom{0}1.8 & \phantom{0}1.8 & \phantom{0}2.5 & \phantom{0}2.4 & \phantom{0}3.5 & \phantom{0}2.3 &
		35.8 & 31.4 & \phantom{0}8.4 & \phantom{0}5.6 & 10.2 & \phantom{0}7.4 &
		21.5 & 19.4 & \phantom{0}7.1 & \phantom{0}9.5 & \phantom{0}2.7 & \phantom{0}2.5 \\

		TF$\times$IDF &
		16.7 & 16.9 & 12.1 & 11.4 & 17.7 & 12.7 &
		\textbf{36.5} & \textbf{34.4} & \phantom{0}9.3 & 10.1 & 11.6 & 12.3 &
		23.3 & 21.6 & 16.9 & 15.8 & \phantom{0}9.6 & \phantom{0}9.4 \\

		\midrule

		PositionRank &
		\phantom{0}4.9 & \phantom{0}4.6 & \phantom{0}5.7 & \phantom{0}4.9 & \phantom{0}6.8 & \phantom{0}4.1 &
		34.2 & 32.2 & 11.6\da & \phantom{0}8.4 & 14.1\da & 11.2 &
		28.6\da & \textbf{28.0}\da & 13.4 & 12.7 & \phantom{0}8.5 & \phantom{0}6.6 \\

		MPRank &
		15.8 & 15.0 & 11.6 & 11.0 & 14.3 & 10.6 &
		30.5 & 29.0 & 10.8\da & 10.4 & 13.6\da & 13.3\da &
		25.6 & 24.9\da & \textbf{18.2} & \textbf{17.0} & 11.2\da & 10.1\da \\

		EmbedRank &
		\phantom{0}3.7 & \phantom{0}3.2 & \phantom{0}2.1 & \phantom{0}2.1 & \phantom{0}2.5 & \phantom{0}2.0 &
		35.6 & 32.5 & 10.7\da & \phantom{0}7.7 & 12.4 & 10.0 &
		\textbf{29.5}\da & 27.5\da & 12.4 & 12.4 & \phantom{0}4.0 & \phantom{0}3.3 \\

		\midrule

		Kea &
		18.6\da & 18.6\da & 14.2\da & 13.3 & 19.5\da & \textbf{14.7}\da &
		34.5 & 33.2 & 11.0\da & 10.9\da & 14.0\da & 13.8\da &
		26.5\da & 24.5\da & 17.3 & 16.7 & 11.0\da & 10.8\da \\

		CopyRNN &
		\textbf{24.2}\da & \textbf{25.4}\da & \textbf{24.4}\da & \textbf{26.3}\da & \textbf{20.3}\da & 13.8 &
		28.2 & 26.4 & \textbf{22.2}\da & \textbf{24.9}\da & \textbf{25.4}\da & \textbf{28.7}\da &
	    10.5 & \phantom{0}7.2 & \phantom{0}8.4 & \phantom{0}4.2 & \textbf{39.3}\da & \textbf{50.9}\da \\

		CorrRNN &
		20.8\da & 19.4\da & 21.1\da & 20.5\da & 19.4 & 10.9 &
		27.9 & 23.6 & 19.9\da & 20.3\da & 21.8\da & 22.7 &
        10.5 & \phantom{0}6.5 & \phantom{0}7.8 & \phantom{0}3.2 & 20.5\da & 20.3\da \\
        

        
        
        \bottomrule
    \end{tabular}
    }
    \caption{Performance of keyphrase extraction models.
    $^\dagger$~indicates significance over the baselines.}
    \label{tab:results}
\end{table*}

\subsection{Parameter settings}

We pre-process all the texts using the Stanford CoreNLP suite~\cite{manning-EtAl:2014:P14-5} for tokenization, sentence splitting and part-of-speech (POS) tagging.
All non-neural models operate on a set of keyphrase candidates, extracted from the input document.
Selecting appropriate candidates is particularly important since it determines the upper bound on recall, and the amount of irrelevant candidates that models will have to deal with.
For a fair and meaningful comparison, we use the same candidate selection heuristic across models.
We follow the recommendation by~\citet{10.1007/978-3-642-54906-9_14} and select the sequences of adjacent nouns with one or more preceding adjectives of length up to five words.
Candidates are further filtered by removing those shorter than 3 characters or containing non-alphanumeric symbols.

We implemented the neural models in PyTorch~\cite{paszke2017automatic} using AllenNLP~\cite{gardner-etal-2018-allennlp}, and the non-neural models using the \texttt{pke} toolkit~\cite{boudin:2016:COLINGDEMO}.
As neural models require large amounts of annotated data to be trained, we trained our models on the KP20k dataset for both scientific papers and abstracts, and on KPTimes for news texts.
%
%
We compute Document Frequency (DF) counts and learn Kea models on training sets.
For datasets without training splits, we apply a leave-one-out cross-validation procedure on the test sets for calculating DF counts and training models.
We use the optimal parameters suggested by the authors for each model, and leverage pre-trained sentence embeddings\footnote{\url{https://github.com/epfml/sent2vec}} for EmbedRank. 
We also found out that the training set of KP20k contains a non-negligible number of documents from the test sets of other datasets.
We removed those documents prior to training.

\subsection{Evaluation metrics}

Although there is no consensus as to which metric is the most reliable for keyphrase extraction, a popular evaluation strategy is to compare the top $k$ extracted keyphrases against the gold standard.
We adopt this strategy and report the f-measure at the top 10 extracted keyphrases.
%
In previous work, we often see differences in how gold standards are handled during evaluation.
%
For example, some studies evaluate their models on the present and missing portions of the gold standard separately~\cite[\textit{inter alia}]{P17-1054,ye-wang:2018:EMNLP,chen-EtAl:2018:EMNLP9}, whereas other work use the entire gold standard~\cite[\textit{inter alia}]{florescu-caragea:2017:Long,boudin:2018:N18-2}.
We chose the latter because recent models, in addition to extracting keyphrases from the content of the document, are able to generate missing keyphrases.
%
%
%
Following common practice, gold standard and output keyphrases are stemmed to reduce the number of mismatches.
One issue with the f-measure is that the ranks of the correct keyphrases are not taken into account.
To evaluate the overall ranking performance of the models, we also report the Mean Average Precision (MAP) scores of the ranked lists of keyphrases.
We use the Student’s paired t-test to assess statistical significance at the $0.05$ level.

\subsection{Replicability of results}

In Table~\ref{tab:replicability}, we compare the results of our re-implementations against those reported in the original papers.
We note that all models show comparable results.
We observe the largest differences with original scores for CopyRNN ($+2$) and CorrRNN ($-4.3$) that can be easily explained by minor differences in training parameters.

\begin{table}[!htb]
    \centering
    \begin{tabular}{lrcc}
    \toprule
    \textbf{Model} & \textbf{Dataset (metric)} & \textbf{Orig.} & \textbf{Ours} \\
    \midrule
        PositionRank & \small{WWW (F$@$8)}       & 12.3 & 11.7 \\
        MPRank & \small{SemEval-2010 (F$@$10)}   & 14.5 & 14.3 \\
        EmbedRank & \small{Inspec (F$@$10)}   & 37.1 & 35.6 \\ 
        CopyRNN & \small{KP20k (F$@$10 on present)} & 26.2 & 28.2 \\
        CorrRNN & \small{ACM (F$@$10 on present)} & 27.8 & 23.5 \\ 
    \bottomrule
    \end{tabular}
    \caption{Original vs. re-implementation scores.}
    \label{tab:replicability}
\end{table}


\section{Results}
\label{sec:results}

Results are presented in Table~\ref{tab:results}.
First of all, we notice that no model significantly outperforms the baselines on all datasets.
This is rather surprising, as one would expect that neural models would be consistently better than a simple TF$\times$IDF model for example.
Rather, we see that the TF$\times$IDF baseline is very competitive on long documents, while the FirstPhrases baseline performs remarkably well, especially on news texts.
Still, overall, CopyRNN achieves the best performance with, in the case of KPTimes, MAP scores exceeding 50\%.
When we look at only unsupervised models, MPRank achieves the best results across datasets.
Also, it comes as no surprise that Kea exhibits strong performance across datasets because it combines two effective features, as demonstrated by the results of the TF$\times$IDF and FirstPhrases baselines.
Conversely, despite the addition of mechanisms for promoting diversity in the output, CorrRNN is almost always outperformed by CopyRNN, suggesting that the added correlation constraints are not effective at filtering out spurious keyphrases.

In light of the above, we can now answer the following question: ``\textit{How much progress have we made since early models?}''.
It is clear that neural-based models are the new state-of-the-art for keyphrase extraction, achieving F@10 scores up to three times that of previous models.
That being said, CopyRNN, which is the best overall model, fails to consistently outperform the baselines on all datasets.
One reason for that is the limited generalization ability of neural-based models~\cite{P17-1054,chen-EtAl:2018:EMNLP9,gallina-etal-2019-kptimes}, which means that their performance degrades on documents that differ from the ones encountered during training.
This is besides confirmed by the extremely low performance of these models on DUC-2001 and KPCrowd.
Much more work needs to be done in tackling this issue if neural models are to substitute for older supervised models.
Perhaps most disappointing is the fact that state-of-the-art unsupervised models are still challenged by the TF$\times$IDF baseline.
Here, we suspect the reasons are twofold.
First, the models we have investigated do not use in-domain data which may not only limit their performance, but also, as in the case of EmbedRank that uses out-of-domain (Wikipedia) data, be detrimental to their performance.
Second, unlike neural generative models, they are not able to produce keyphrases that do not occur in the source document, further limiting their potential effectiveness.

As outlined in \S\ref{sec:datasets}, gold standards provided by lay annotators, such as authors and readers, exhibit strong inconsistency issues.
One might therefore wonder \textit{``What is the impact of non-expert annotations on training and evaluating keyphrase extraction models?''}.
Intuitively, models evaluated against these annotations are likely to receive lower scores because they make training more difficult (that is, assigning different keyphrases to documents about the same topic may confuse the model) while increasing the number of false negatives during evaluation.
This is exactly what we observe in Table~\ref{tab:results} where the best scores for Inspec and KPTimes, whose gold standards are provided by professional indexers, are higher in magnitude than those of the other datasets.
Precisely quantifying how much impact lay annotations have on performance is no easy task as it implies a double-annotation process by both expert and non-expert annotators.
Luckily enough, a small sample of documents from Inspec are also found in KP20k, allowing us to compare the performance of keyphrases models between both annotation types.
Results are shown in Table~\ref{tab:lay_vs_expert}.
First, we see that overall performance is nearly cut in half when evaluating against author-provided gold standard, suggesting that reported scores in previous studies are arguably underestimated.
Second, neural models again do not show their superiority against indexer-assigned keyphrases, which advocates the need for more experiments on datasets that include expert annotations.

\begin{table}[!htb]
    \centering
    \begin{tabular}{r c@{\hspace*{2mm}}c  c@{\hspace*{2mm}}c}
            ~ & 
            \multicolumn{2}{c}{$\text{F}@10$} &
            \multicolumn{2}{c}{MAP} \\
            
            \cmidrule(lr){2-3} \cmidrule(lr){4-5} \\ [-1.5em]
            
            \textbf{Model} & \small{\textit{I}} & \small{\textit{A}} & \small{\textit{I}} & \small{\textit{A}} \\[-.2em]
            \midrule
            FirstPhrases  & 25.8 & 13.7 & 26.1 & 13.2 \\
            TextRank      & 33.4 & 12.2 & 29.6 & \phantom{0}9.3 \\
            TF$\times$IDF  & \textbf{34.6} & 14.2 & \textbf{33.3} & 16.1 \\
            \midrule
            PositionRank  & 32.9 & 15.9 & 31.0 & 13.0 \\
            MPRank        & 26.4 & 13.8 & 27.6 & 13.6 \\
            EmbedRank     & 34.3 & 15.3 & 31.3 & 11.5 \\
            \midrule
            Kea           & 32.5 & 15.2 & 31.9 & 15.9 \\
            CopyRNN       & 33.7 & \textbf{28.9}\dda & 29.8 & \textbf{33.8}\dda \\
            CorrRNN       & 28.6 & 25.3 & 24.2 & 28.2 \\
            
            \midrule
            
            Avg. & 31.3 & 17.2 & 29.4 & 17.2 \\
            
            \bottomrule
    \end{tabular}
    \caption{Results on a subset of 55 documents from Inspec for indexer (\textit{I}) and author (\textit{A}) gold annotations. \dda~indicates significance over every other model.}
    \label{tab:lay_vs_expert}
\end{table}

\begin{figure*}[htbp]
  \includegraphics[width=\linewidth]{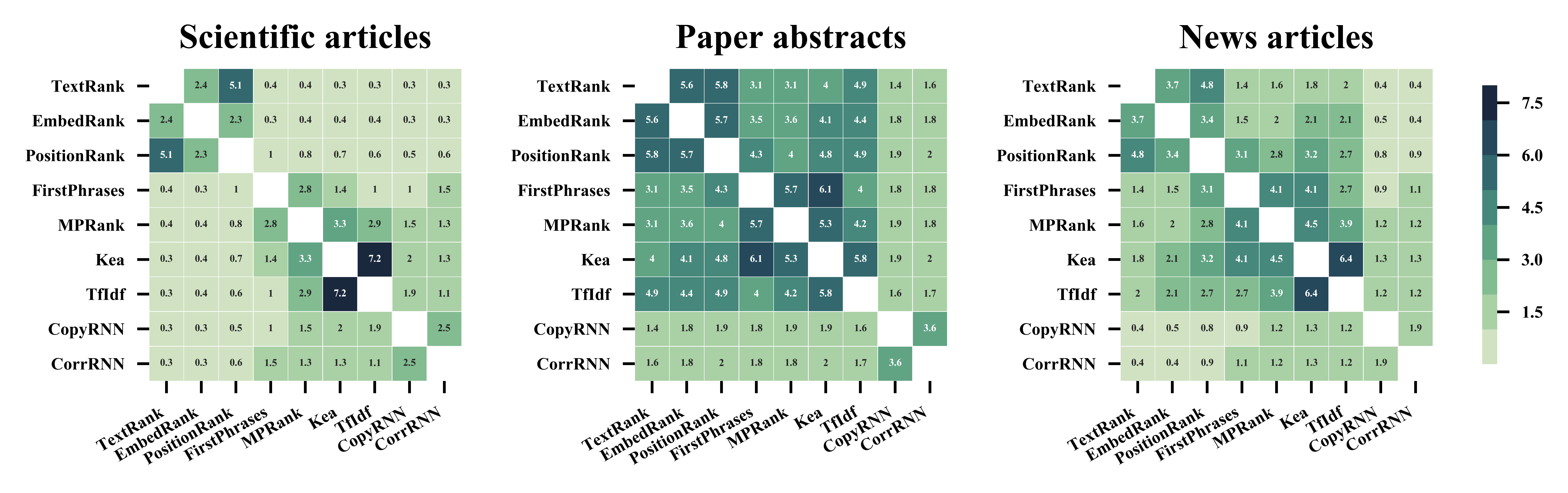}
  \caption{Average number of keyphrases in common between model outputs.}
  \label{fig:intersect}
\end{figure*}

The third question we want to address in this study is ``\textit{Which baselines and benchmark datasets should be included in future work for a better understanding of the pros and cons of a newly proposed model?}''.
Having strong baselines to compare with is of utmost importance, and our results give an indication of which model is relevant.
When properly trained, neural models drastically outperform all other models and represent the state-of-the-art.
Since CopyRNN achieve the best results, it should be included in future work for comparison.
In an unsupervised setting, or in a data-sparse scenario where neural models can not be applied, the picture is less clear.
To help us understand which model is worth investigating, we conducted an additional set of experiments aimed at comparing the outputs from all models in a pairwise manner.
The motivation behind these experiments is that including multiple models that behave similarly is of limited interest.
Similarities between model outputs, viewed in terms of the number of keyphrases in common, are graphed as a heatmap in Figure~\ref{fig:intersect}.
Overall, we observe different patterns for each source of documents.
The shorter the document is, the more similar outputs are, which is mostly due to a smaller search space (that is, a smaller number of keyphrase candidates).
We note that the three best unsupervised models, namely FirstPhrases, MPRank and TF$\times$IDF, generate very similar keyphrases (up to 42\% identical).
Considering this, and given their reported performances (Table~\ref{tab:results}), we argue that TF$\times$IDF (or KEA if seed training data is available) should be considered as strong unsupervised baseline in subsequent work.

These recommendations of baselines also affect the choice of which benchmark datasets one has to use.
As neural models are data-hungry, KP20k and KPTimes are the default options for paper abstracts and news articles.
For scientific articles, we recommend using SemEval for two reasons: 1) it is widely used by existing studies; and 2) it provides a double-annotated gold standard (author- and reader-assigned keyphrases) that alleviates annotation inconsistencies to some extent.

Our experiments highlight several issues in evaluating keyphrase extraction models with existing benchmark datasets.
Another way of assessing the effectiveness of these models would be to explore their impact on other tasks as an extrinsic evaluation.
To the best of our knowledge, there is no previously published research on that matter despite many downstream tasks that already benefit from keyphrase information such as article recommendation~\cite{DBLP:conf/jcdl/CollinsB19} or browsing interfaces~\cite{Gutwin:1999:IBD:338985.338996} in digital libraries.
This points to an interesting future direction that allows for a deeper understanding of the limitations of current models.

\section{Conclusion}

This paper presents a large scale evaluation of keyphrase extraction models conducted on multiple benchmark datasets from different sources and domains.
Results indicate that keyphrase extraction is still an open research question, with state-of-the-art neural-based models still challenged by simple baselines on some datasets.
We hope that this work will serve as a point of departure for more rigorous analysis and evaluation of proposed keyphrase extraction models.
We provide all the code and data on a public repository\footnote{Link to the repository will appear here after the review period.}, as well as a public leaderboard to facilitate the comparison between models.

\bibliographystyle{ACM-Reference-Format}
\bibliography{biblio}


\begin{thebibliography}{50}


\ifx \showCODEN    \undefined \def \showCODEN     #1{\unskip}     \fi
\ifx \showDOI      \undefined \def \showDOI       #1{#1}\fi
\ifx \showISBNx    \undefined \def \showISBNx     #1{\unskip}     \fi
\ifx \showISBNxiii \undefined \def \showISBNxiii  #1{\unskip}     \fi
\ifx \showISSN     \undefined \def \showISSN      #1{\unskip}     \fi
\ifx \showLCCN     \undefined \def \showLCCN      #1{\unskip}     \fi
\ifx \shownote     \undefined \def \shownote      #1{#1}          \fi
\ifx \showarticletitle \undefined \def \showarticletitle #1{#1}   \fi
\ifx \showURL      \undefined \def \showURL       {\relax}        \fi
\providecommand\bibfield[2]{#2}
\providecommand\bibinfo[2]{#2}
\providecommand\natexlab[1]{#1}
\providecommand\showeprint[2][]{arXiv:#2}

\bibitem[\protect\citeauthoryear{Augenstein, Das, Riedel, Vikraman, and
  McCallum}{Augenstein et~al\mbox{.}}{2017}]%
        {augenstein-EtAl:2017:SemEval}
\bibfield{author}{\bibinfo{person}{Isabelle Augenstein},
  \bibinfo{person}{Mrinal Das}, \bibinfo{person}{Sebastian Riedel},
  \bibinfo{person}{Lakshmi Vikraman}, {and} \bibinfo{person}{Andrew McCallum}.}
  \bibinfo{year}{2017}\natexlab{}.
\newblock \showarticletitle{SemEval 2017 Task 10: ScienceIE - Extracting
  Keyphrases and Relations from Scientific Publications}. In
  \bibinfo{booktitle}{\emph{Proceedings of the 11th International Workshop on
  Semantic Evaluation (SemEval-2017)}}. \bibinfo{publisher}{Association for
  Computational Linguistics}, \bibinfo{address}{Vancouver, Canada},
  \bibinfo{pages}{546--555}.
\newblock
\urldef\tempurl%
\url{http://www.aclweb.org/anthology/S17-2091}
\showURL{%
\tempurl}


\bibitem[\protect\citeauthoryear{Bahdanau, Cho, and Bengio}{Bahdanau
  et~al\mbox{.}}{2014}]%
        {bahdanau2014neural}
\bibfield{author}{\bibinfo{person}{Dzmitry Bahdanau},
  \bibinfo{person}{Kyunghyun Cho}, {and} \bibinfo{person}{Yoshua Bengio}.}
  \bibinfo{year}{2014}\natexlab{}.
\newblock \showarticletitle{Neural machine translation by jointly learning to
  align and translate}.
\newblock \bibinfo{journal}{\emph{arXiv preprint arXiv:1409.0473}}
  (\bibinfo{year}{2014}).
\newblock


\bibitem[\protect\citeauthoryear{Barker, Veal, and Wyatt}{Barker
  et~al\mbox{.}}{1972}]%
        {barker1972comparative}
\bibfield{author}{\bibinfo{person}{Frances~H Barker},
  \bibinfo{person}{Douglas~C Veal}, {and} \bibinfo{person}{Barry~K Wyatt}.}
  \bibinfo{year}{1972}\natexlab{}.
\newblock \showarticletitle{Comparative efficiency of searching titles,
  abstracts, and index terms in a free-text data base}.
\newblock \bibinfo{journal}{\emph{Journal of Documentation}}
  \bibinfo{volume}{28}, \bibinfo{number}{1} (\bibinfo{year}{1972}),
  \bibinfo{pages}{22--36}.
\newblock


\bibitem[\protect\citeauthoryear{Bennani-Smires, Musat, Hossmann, Baeriswyl,
  and Jaggi}{Bennani-Smires et~al\mbox{.}}{2018}]%
        {bennanismires-EtAl:2018:K18-1}
\bibfield{author}{\bibinfo{person}{Kamil Bennani-Smires},
  \bibinfo{person}{Claudiu Musat}, \bibinfo{person}{Andreea Hossmann},
  \bibinfo{person}{Michael Baeriswyl}, {and} \bibinfo{person}{Martin Jaggi}.}
  \bibinfo{year}{2018}\natexlab{}.
\newblock \showarticletitle{Simple Unsupervised Keyphrase Extraction using
  Sentence Embeddings}. In \bibinfo{booktitle}{\emph{Proceedings of the 22nd
  Conference on Computational Natural Language Learning}}.
  \bibinfo{publisher}{Association for Computational Linguistics},
  \bibinfo{address}{Brussels, Belgium}, \bibinfo{pages}{221--229}.
\newblock
\urldef\tempurl%
\url{http://www.aclweb.org/anthology/K18-1022}
\showURL{%
\tempurl}


\bibitem[\protect\citeauthoryear{Berend}{Berend}{2011}]%
        {berend:2011:IJCNLP-2011}
\bibfield{author}{\bibinfo{person}{G\'{a}bor Berend}.}
  \bibinfo{year}{2011}\natexlab{}.
\newblock \showarticletitle{Opinion Expression Mining by Exploiting Keyphrase
  Extraction}. In \bibinfo{booktitle}{\emph{Proceedings of 5th International
  Joint Conference on Natural Language Processing}}. \bibinfo{publisher}{Asian
  Federation of Natural Language Processing}, \bibinfo{address}{Chiang Mai,
  Thailand}, \bibinfo{pages}{1162--1170}.
\newblock
\urldef\tempurl%
\url{http://www.aclweb.org/anthology/I11-1130}
\showURL{%
\tempurl}


\bibitem[\protect\citeauthoryear{Boudin}{Boudin}{2016}]%
        {boudin:2016:COLINGDEMO}
\bibfield{author}{\bibinfo{person}{Florian Boudin}.}
  \bibinfo{year}{2016}\natexlab{}.
\newblock \showarticletitle{pke: an open source python-based keyphrase
  extraction toolkit}. In \bibinfo{booktitle}{\emph{Proceedings of COLING 2016,
  the 26th International Conference on Computational Linguistics: System
  Demonstrations}}. \bibinfo{publisher}{The COLING 2016 Organizing Committee},
  \bibinfo{address}{Osaka, Japan}, \bibinfo{pages}{69--73}.
\newblock
\urldef\tempurl%
\url{http://aclweb.org/anthology/C16-2015}
\showURL{%
\tempurl}


\bibitem[\protect\citeauthoryear{Boudin}{Boudin}{2018}]%
        {boudin:2018:N18-2}
\bibfield{author}{\bibinfo{person}{Florian Boudin}.}
  \bibinfo{year}{2018}\natexlab{}.
\newblock \showarticletitle{Unsupervised Keyphrase Extraction with Multipartite
  Graphs}. In \bibinfo{booktitle}{\emph{Proceedings of the 2018 Conference of
  the North American Chapter of the Association for Computational Linguistics:
  Human Language Technologies, Volume 2 (Short Papers)}}.
  \bibinfo{publisher}{Association for Computational Linguistics},
  \bibinfo{address}{New Orleans, Louisiana}, \bibinfo{pages}{667--672}.
\newblock
\urldef\tempurl%
\url{http://www.aclweb.org/anthology/N18-2105}
\showURL{%
\tempurl}


\bibitem[\protect\citeauthoryear{Boudin, Mougard, and Cram}{Boudin
  et~al\mbox{.}}{2016}]%
        {boudin-mougard-cram:2016:WNUT}
\bibfield{author}{\bibinfo{person}{Florian Boudin}, \bibinfo{person}{Hugo
  Mougard}, {and} \bibinfo{person}{Damien Cram}.}
  \bibinfo{year}{2016}\natexlab{}.
\newblock \showarticletitle{How Document Pre-processing affects Keyphrase
  Extraction Performance}. In \bibinfo{booktitle}{\emph{Proceedings of the 2nd
  Workshop on Noisy User-generated Text (WNUT)}}. \bibinfo{publisher}{The
  COLING 2016 Organizing Committee}, \bibinfo{address}{Osaka, Japan},
  \bibinfo{pages}{121--128}.
\newblock
\urldef\tempurl%
\url{http://aclweb.org/anthology/W16-3917}
\showURL{%
\tempurl}


\bibitem[\protect\citeauthoryear{Caragea, Bulgarov, Godea, and
  Das~Gollapalli}{Caragea et~al\mbox{.}}{2014}]%
        {caragea-EtAl:2014:EMNLP2014}
\bibfield{author}{\bibinfo{person}{Cornelia Caragea},
  \bibinfo{person}{Florin~Adrian Bulgarov}, \bibinfo{person}{Andreea Godea},
  {and} \bibinfo{person}{Sujatha Das~Gollapalli}.}
  \bibinfo{year}{2014}\natexlab{}.
\newblock \showarticletitle{Citation-Enhanced Keyphrase Extraction from
  Research Papers: A Supervised Approach}. In
  \bibinfo{booktitle}{\emph{Proceedings of the 2014 Conference on Empirical
  Methods in Natural Language Processing (EMNLP)}}.
  \bibinfo{publisher}{Association for Computational Linguistics},
  \bibinfo{address}{Doha, Qatar}, \bibinfo{pages}{1435--1446}.
\newblock
\urldef\tempurl%
\url{http://www.aclweb.org/anthology/D14-1150}
\showURL{%
\tempurl}


\bibitem[\protect\citeauthoryear{Chen, Zhang, Wu, Yan, and Li}{Chen
  et~al\mbox{.}}{2018}]%
        {chen-EtAl:2018:EMNLP9}
\bibfield{author}{\bibinfo{person}{Jun Chen}, \bibinfo{person}{Xiaoming Zhang},
  \bibinfo{person}{Yu Wu}, \bibinfo{person}{Zhao Yan}, {and}
  \bibinfo{person}{Zhoujun Li}.} \bibinfo{year}{2018}\natexlab{}.
\newblock \showarticletitle{Keyphrase Generation with Correlation Constraints}.
  In \bibinfo{booktitle}{\emph{Proceedings of the 2018 Conference on Empirical
  Methods in Natural Language Processing}}. \bibinfo{publisher}{Association for
  Computational Linguistics}, \bibinfo{address}{Brussels, Belgium},
  \bibinfo{pages}{4057--4066}.
\newblock
\urldef\tempurl%
\url{http://www.aclweb.org/anthology/D18-1439}
\showURL{%
\tempurl}


\bibitem[\protect\citeauthoryear{Cho, van Merrienboer, Gulcehre, Bahdanau,
  Bougares, Schwenk, and Bengio}{Cho et~al\mbox{.}}{2014}]%
        {cho-EtAl:2014:EMNLP2014}
\bibfield{author}{\bibinfo{person}{Kyunghyun Cho}, \bibinfo{person}{Bart van
  Merrienboer}, \bibinfo{person}{Caglar Gulcehre}, \bibinfo{person}{Dzmitry
  Bahdanau}, \bibinfo{person}{Fethi Bougares}, \bibinfo{person}{Holger
  Schwenk}, {and} \bibinfo{person}{Yoshua Bengio}.}
  \bibinfo{year}{2014}\natexlab{}.
\newblock \showarticletitle{Learning Phrase Representations using RNN
  Encoder--Decoder for Statistical Machine Translation}. In
  \bibinfo{booktitle}{\emph{Proceedings of the 2014 Conference on Empirical
  Methods in Natural Language Processing (EMNLP)}}.
  \bibinfo{publisher}{Association for Computational Linguistics},
  \bibinfo{address}{Doha, Qatar}, \bibinfo{pages}{1724--1734}.
\newblock
\urldef\tempurl%
\url{http://www.aclweb.org/anthology/D14-1179}
\showURL{%
\tempurl}


\bibitem[\protect\citeauthoryear{Collins and Beel}{Collins and Beel}{2019}]%
        {DBLP:conf/jcdl/CollinsB19}
\bibfield{author}{\bibinfo{person}{Andrew Collins} {and}
  \bibinfo{person}{J{\"{o}}ran Beel}.} \bibinfo{year}{2019}\natexlab{}.
\newblock \showarticletitle{Document Embeddings vs. Keyphrases vs. Terms for
  Recommender Systems: {A} Large-Scale Online Evaluation}. In
  \bibinfo{booktitle}{\emph{19th {ACM/IEEE} Joint Conference on Digital
  Libraries, {JCDL} 2019, Champaign, IL, USA, June 2-6, 2019}}.
  \bibinfo{pages}{130--133}.
\newblock
\urldef\tempurl%
\url{https://doi.org/10.1109/JCDL.2019.00027}
\showDOI{\tempurl}


\bibitem[\protect\citeauthoryear{Evans and Zhai}{Evans and Zhai}{1996}]%
        {evans-zhai:1996:ACL}
\bibfield{author}{\bibinfo{person}{David~A. Evans} {and}
  \bibinfo{person}{Chengxiang Zhai}.} \bibinfo{year}{1996}\natexlab{}.
\newblock \showarticletitle{Noun Phrase Analysis in Large Unrestricted Text for
  Information Retrieval}. In \bibinfo{booktitle}{\emph{Proceedings of the 34th
  Annual Meeting of the Association for Computational Linguistics}}.
  \bibinfo{publisher}{Association for Computational Linguistics},
  \bibinfo{address}{Santa Cruz, California, USA}, \bibinfo{pages}{17--24}.
\newblock
\urldef\tempurl%
\url{https://doi.org/10.3115/981863.981866}
\showDOI{\tempurl}


\bibitem[\protect\citeauthoryear{Florescu and Caragea}{Florescu and
  Caragea}{2017}]%
        {florescu-caragea:2017:Long}
\bibfield{author}{\bibinfo{person}{Corina Florescu} {and}
  \bibinfo{person}{Cornelia Caragea}.} \bibinfo{year}{2017}\natexlab{}.
\newblock \showarticletitle{PositionRank: An Unsupervised Approach to Keyphrase
  Extraction from Scholarly Documents}. In
  \bibinfo{booktitle}{\emph{Proceedings of the 55th Annual Meeting of the
  Association for Computational Linguistics (Volume 1: Long Papers)}}.
  \bibinfo{publisher}{Association for Computational Linguistics},
  \bibinfo{address}{Vancouver, Canada}, \bibinfo{pages}{1105--1115}.
\newblock
\urldef\tempurl%
\url{http://aclweb.org/anthology/P17-1102}
\showURL{%
\tempurl}


\bibitem[\protect\citeauthoryear{Gallina, Boudin, and Daille}{Gallina
  et~al\mbox{.}}{2019}]%
        {gallina-etal-2019-kptimes}
\bibfield{author}{\bibinfo{person}{Ygor Gallina}, \bibinfo{person}{Florian
  Boudin}, {and} \bibinfo{person}{Beatrice Daille}.}
  \bibinfo{year}{2019}\natexlab{}.
\newblock \showarticletitle{{KPT}imes: A Large-Scale Dataset for Keyphrase
  Generation on News Documents}. In \bibinfo{booktitle}{\emph{Proceedings of
  the 12th International Conference on Natural Language Generation}}.
  \bibinfo{publisher}{Association for Computational Linguistics},
  \bibinfo{address}{Tokyo, Japan}, \bibinfo{pages}{130--135}.
\newblock
\urldef\tempurl%
\url{https://doi.org/10.18653/v1/W19-8617}
\showDOI{\tempurl}


\bibitem[\protect\citeauthoryear{Gardner, Grus, Neumann, Tafjord, Dasigi, Liu,
  Peters, Schmitz, and Zettlemoyer}{Gardner et~al\mbox{.}}{2018}]%
        {gardner-etal-2018-allennlp}
\bibfield{author}{\bibinfo{person}{Matt Gardner}, \bibinfo{person}{Joel Grus},
  \bibinfo{person}{Mark Neumann}, \bibinfo{person}{Oyvind Tafjord},
  \bibinfo{person}{Pradeep Dasigi}, \bibinfo{person}{Nelson~F. Liu},
  \bibinfo{person}{Matthew Peters}, \bibinfo{person}{Michael Schmitz}, {and}
  \bibinfo{person}{Luke Zettlemoyer}.} \bibinfo{year}{2018}\natexlab{}.
\newblock \showarticletitle{{A}llen{NLP}: A Deep Semantic Natural Language
  Processing Platform}. In \bibinfo{booktitle}{\emph{Proceedings of Workshop
  for {NLP} Open Source Software ({NLP}-{OSS})}}.
  \bibinfo{publisher}{Association for Computational Linguistics},
  \bibinfo{address}{Melbourne, Australia}, \bibinfo{pages}{1--6}.
\newblock
\urldef\tempurl%
\url{https://doi.org/10.18653/v1/W18-2501}
\showDOI{\tempurl}


\bibitem[\protect\citeauthoryear{Goldstein and Carbonell}{Goldstein and
  Carbonell}{1998}]%
        {goldstein-carbonell:1998:TIPSTER98}
\bibfield{author}{\bibinfo{person}{Jade Goldstein} {and} \bibinfo{person}{Jaime
  Carbonell}.} \bibinfo{year}{1998}\natexlab{}.
\newblock \showarticletitle{SUMMARIZATION: (1) USING MMR FOR DIVERSITY- BASED
  RERANKING AND (2) EVALUATING SUMMARIES}. In
  \bibinfo{booktitle}{\emph{Proceedings of the TIPSTER Text Program: Phase
  III}}. \bibinfo{publisher}{Association for Computational Linguistics},
  \bibinfo{address}{Baltimore, Maryland, USA}, \bibinfo{pages}{181--195}.
\newblock
\urldef\tempurl%
\url{https://doi.org/10.3115/1119089.1119120}
\showDOI{\tempurl}


\bibitem[\protect\citeauthoryear{Gollapalli, Caragea, Li, and Giles}{Gollapalli
  et~al\mbox{.}}{2015}]%
        {Keyphrase:2015}
\bibfield{editor}{\bibinfo{person}{Sujatha~Das Gollapalli},
  \bibinfo{person}{Cornelia Caragea}, \bibinfo{person}{Xiaoli Li}, {and}
  \bibinfo{person}{C.~Lee Giles}} (Eds.). \bibinfo{year}{2015}\natexlab{}.
\newblock \bibinfo{booktitle}{\emph{Proceedings of the ACL 2015 Workshop on
  Novel Computational Approaches to Keyphrase Extraction}}.
\newblock \bibinfo{publisher}{Association for Computational Linguistics},
  \bibinfo{address}{Beijing, China}.
\newblock
\urldef\tempurl%
\url{http://www.aclweb.org/anthology/W15-36}
\showURL{%
\tempurl}


\bibitem[\protect\citeauthoryear{Gu, Lu, Li, and Li}{Gu et~al\mbox{.}}{2016}]%
        {gu-EtAl:2016:P16-1}
\bibfield{author}{\bibinfo{person}{Jiatao Gu}, \bibinfo{person}{Zhengdong Lu},
  \bibinfo{person}{Hang Li}, {and} \bibinfo{person}{Victor~O.K. Li}.}
  \bibinfo{year}{2016}\natexlab{}.
\newblock \showarticletitle{Incorporating Copying Mechanism in
  Sequence-to-Sequence Learning}. In \bibinfo{booktitle}{\emph{Proceedings of
  the 54th Annual Meeting of the Association for Computational Linguistics
  (Volume 1: Long Papers)}}. \bibinfo{publisher}{Association for Computational
  Linguistics}, \bibinfo{address}{Berlin, Germany},
  \bibinfo{pages}{1631--1640}.
\newblock
\urldef\tempurl%
\url{http://www.aclweb.org/anthology/P16-1154}
\showURL{%
\tempurl}


\bibitem[\protect\citeauthoryear{Gutwin, Paynter, Witten, Nevill-Manning, and
  Frank}{Gutwin et~al\mbox{.}}{1999}]%
        {Gutwin:1999:IBD:338985.338996}
\bibfield{author}{\bibinfo{person}{Carl Gutwin}, \bibinfo{person}{Gordon
  Paynter}, \bibinfo{person}{Ian Witten}, \bibinfo{person}{Craig
  Nevill-Manning}, {and} \bibinfo{person}{Eibe Frank}.}
  \bibinfo{year}{1999}\natexlab{}.
\newblock \showarticletitle{Improving Browsing in Digital Libraries with
  Keyphrase Indexes}.
\newblock \bibinfo{journal}{\emph{Decis. Support Syst.}} \bibinfo{volume}{27},
  \bibinfo{number}{1-2} (\bibinfo{date}{Nov.} \bibinfo{year}{1999}),
  \bibinfo{pages}{81--104}.
\newblock
\showISSN{0167-9236}
\urldef\tempurl%
\url{https://doi.org/10.1016/S0167-9236(99)00038-X}
\showDOI{\tempurl}


\bibitem[\protect\citeauthoryear{Hasan and Ng}{Hasan and Ng}{2014}]%
        {hasan-ng:2014:P14-1}
\bibfield{author}{\bibinfo{person}{Kazi~Saidul Hasan} {and}
  \bibinfo{person}{Vincent Ng}.} \bibinfo{year}{2014}\natexlab{}.
\newblock \showarticletitle{Automatic Keyphrase Extraction: A Survey of the
  State of the Art}. In \bibinfo{booktitle}{\emph{Proceedings of the 52nd
  Annual Meeting of the Association for Computational Linguistics (Volume 1:
  Long Papers)}}. \bibinfo{publisher}{Association for Computational
  Linguistics}, \bibinfo{address}{Baltimore, Maryland},
  \bibinfo{pages}{1262--1273}.
\newblock
\urldef\tempurl%
\url{http://www.aclweb.org/anthology/P14-1119}
\showURL{%
\tempurl}


\bibitem[\protect\citeauthoryear{Hulth}{Hulth}{2003}]%
        {Hulth:2003:EMNLP}
\bibfield{author}{\bibinfo{person}{Anette Hulth}.}
  \bibinfo{year}{2003}\natexlab{}.
\newblock \showarticletitle{Improved Automatic Keyword Extraction Given More
  Linguistic Knowledge}. In \bibinfo{booktitle}{\emph{Proceedings of the 2003
  Conference on Empirical Methods in Natural Language Processing}},
  \bibfield{editor}{\bibinfo{person}{Michael Collins} {and}
  \bibinfo{person}{Mark Steedman}} (Eds.). \bibinfo{pages}{216--223}.
\newblock
\urldef\tempurl%
\url{http://www.aclweb.org/anthology/W03-1028.pdf}
\showURL{%
\tempurl}


\bibitem[\protect\citeauthoryear{Hulth and Megyesi}{Hulth and Megyesi}{2006}]%
        {hulth-megyesi:2006:COLACL}
\bibfield{author}{\bibinfo{person}{Anette Hulth} {and}
  \bibinfo{person}{Be\'{a}ta~B. Megyesi}.} \bibinfo{year}{2006}\natexlab{}.
\newblock \showarticletitle{A Study on Automatically Extracted Keywords in Text
  Categorization}. In \bibinfo{booktitle}{\emph{Proceedings of the 21st
  International Conference on Computational Linguistics and 44th Annual Meeting
  of the Association for Computational Linguistics}}.
  \bibinfo{publisher}{Association for Computational Linguistics},
  \bibinfo{address}{Sydney, Australia}, \bibinfo{pages}{537--544}.
\newblock
\urldef\tempurl%
\url{https://doi.org/10.3115/1220175.1220243}
\showDOI{\tempurl}


\bibitem[\protect\citeauthoryear{Hussey, Williams, Mitchell, and Field}{Hussey
  et~al\mbox{.}}{2012}]%
        {reading32266}
\bibfield{author}{\bibinfo{person}{Richard Hussey}, \bibinfo{person}{Shirley
  Williams}, \bibinfo{person}{Richard Mitchell}, {and} \bibinfo{person}{Ian
  Field}.} \bibinfo{year}{2012}\natexlab{}.
\newblock \showarticletitle{A comparison of automated keyphrase extraction
  techniques and of automatic evaluation vs. human evaluation}.
\newblock \bibinfo{journal}{\emph{International Journal on Advances in Life
  Sciences}} \bibinfo{volume}{4}, \bibinfo{number}{3 and 4}
  (\bibinfo{year}{2012}), \bibinfo{pages}{136--153}.
\newblock
\urldef\tempurl%
\url{http://centaur.reading.ac.uk/32266/}
\showURL{%
\tempurl}


\bibitem[\protect\citeauthoryear{Kedzie, McKeown, and Daume~III}{Kedzie
  et~al\mbox{.}}{2018}]%
        {kedzie-mckeown-daumeiii:2018:EMNLP}
\bibfield{author}{\bibinfo{person}{Chris Kedzie}, \bibinfo{person}{Kathleen
  McKeown}, {and} \bibinfo{person}{Hal Daume~III}.}
  \bibinfo{year}{2018}\natexlab{}.
\newblock \showarticletitle{Content Selection in Deep Learning Models of
  Summarization}. In \bibinfo{booktitle}{\emph{Proceedings of the 2018
  Conference on Empirical Methods in Natural Language Processing}}.
  \bibinfo{publisher}{Association for Computational Linguistics},
  \bibinfo{address}{Brussels, Belgium}, \bibinfo{pages}{1818--1828}.
\newblock
\urldef\tempurl%
\url{http://www.aclweb.org/anthology/D18-1208}
\showURL{%
\tempurl}


\bibitem[\protect\citeauthoryear{Kim, Medelyan, Kan, and Baldwin}{Kim
  et~al\mbox{.}}{2010}]%
        {kim-EtAl:2010:SemEval}
\bibfield{author}{\bibinfo{person}{Su~Nam Kim}, \bibinfo{person}{Olena
  Medelyan}, \bibinfo{person}{Min-Yen Kan}, {and} \bibinfo{person}{Timothy
  Baldwin}.} \bibinfo{year}{2010}\natexlab{}.
\newblock \showarticletitle{SemEval-2010 Task 5 : Automatic Keyphrase
  Extraction from Scientific Articles}. In
  \bibinfo{booktitle}{\emph{Proceedings of the 5th International Workshop on
  Semantic Evaluation}}. \bibinfo{publisher}{Association for Computational
  Linguistics}, \bibinfo{address}{Uppsala, Sweden}, \bibinfo{pages}{21--26}.
\newblock
\urldef\tempurl%
\url{http://www.aclweb.org/anthology/S10-1004}
\showURL{%
\tempurl}


\bibitem[\protect\citeauthoryear{Krapivin, Autaeu, and Marchese}{Krapivin
  et~al\mbox{.}}{2009}]%
        {krapivin2009large}
\bibfield{author}{\bibinfo{person}{Mikalai Krapivin},
  \bibinfo{person}{Aliaksandr Autaeu}, {and} \bibinfo{person}{Maurizio
  Marchese}.} \bibinfo{year}{2009}\natexlab{}.
\newblock \bibinfo{booktitle}{\emph{Large dataset for keyphrases extraction}}.
\newblock \bibinfo{type}{{T}echnical {R}eport}.
  \bibinfo{institution}{University of Trento}.
\newblock


\bibitem[\protect\citeauthoryear{Litvak and Last}{Litvak and Last}{2008}]%
        {litvak-last:2008:MMIES}
\bibfield{author}{\bibinfo{person}{Marina Litvak} {and} \bibinfo{person}{Mark
  Last}.} \bibinfo{year}{2008}\natexlab{}.
\newblock \showarticletitle{Graph-Based Keyword Extraction for Single-Document
  Summarization}. In \bibinfo{booktitle}{\emph{Coling 2008: Proceedings of the
  workshop Multi-source Multilingual Information Extraction and
  Summarization}}. \bibinfo{publisher}{Coling 2008 Organizing Committee},
  \bibinfo{address}{Manchester, UK}, \bibinfo{pages}{17--24}.
\newblock
\urldef\tempurl%
\url{http://www.aclweb.org/anthology/W08-1404}
\showURL{%
\tempurl}


\bibitem[\protect\citeauthoryear{Luong, Pham, and Manning}{Luong
  et~al\mbox{.}}{2015}]%
        {luong-pham-manning:2015:EMNLP}
\bibfield{author}{\bibinfo{person}{Thang Luong}, \bibinfo{person}{Hieu Pham},
  {and} \bibinfo{person}{Christopher~D. Manning}.}
  \bibinfo{year}{2015}\natexlab{}.
\newblock \showarticletitle{Effective Approaches to Attention-based Neural
  Machine Translation}. In \bibinfo{booktitle}{\emph{Proceedings of the 2015
  Conference on Empirical Methods in Natural Language Processing}}.
  \bibinfo{publisher}{Association for Computational Linguistics},
  \bibinfo{address}{Lisbon, Portugal}, \bibinfo{pages}{1412--1421}.
\newblock
\urldef\tempurl%
\url{http://aclweb.org/anthology/D15-1166}
\showURL{%
\tempurl}


\bibitem[\protect\citeauthoryear{Manning, Surdeanu, Bauer, Finkel, Bethard, and
  McClosky}{Manning et~al\mbox{.}}{2014}]%
        {manning-EtAl:2014:P14-5}
\bibfield{author}{\bibinfo{person}{Christopher~D. Manning},
  \bibinfo{person}{Mihai Surdeanu}, \bibinfo{person}{John Bauer},
  \bibinfo{person}{Jenny Finkel}, \bibinfo{person}{Steven~J. Bethard}, {and}
  \bibinfo{person}{David McClosky}.} \bibinfo{year}{2014}\natexlab{}.
\newblock \showarticletitle{The {Stanford} {CoreNLP} Natural Language
  Processing Toolkit}. In \bibinfo{booktitle}{\emph{Association for
  Computational Linguistics (ACL) System Demonstrations}}.
  \bibinfo{pages}{55--60}.
\newblock
\urldef\tempurl%
\url{http://www.aclweb.org/anthology/P/P14/P14-5010}
\showURL{%
\tempurl}


\bibitem[\protect\citeauthoryear{Mao and Lu}{Mao and Lu}{2017}]%
        {Mao2017}
\bibfield{author}{\bibinfo{person}{Yuqing Mao} {and} \bibinfo{person}{Zhiyong
  Lu}.} \bibinfo{year}{2017}\natexlab{}.
\newblock \showarticletitle{MeSH Now: automatic MeSH indexing at PubMed scale
  via learning to rank}.
\newblock \bibinfo{journal}{\emph{Journal of Biomedical Semantics}}
  \bibinfo{volume}{8}, \bibinfo{number}{1} (\bibinfo{date}{17 Apr}
  \bibinfo{year}{2017}), \bibinfo{pages}{15}.
\newblock
\showISSN{2041-1480}
\urldef\tempurl%
\url{https://doi.org/10.1186/s13326-017-0123-3}
\showDOI{\tempurl}


\bibitem[\protect\citeauthoryear{Marcu}{Marcu}{1997}]%
        {marcu:1997:ACL}
\bibfield{author}{\bibinfo{person}{Daniel Marcu}.}
  \bibinfo{year}{1997}\natexlab{}.
\newblock \showarticletitle{The Rhetorical Parsing of Unrestricted Natural
  Language Texts}. In \bibinfo{booktitle}{\emph{Proceedings of the 35th Annual
  Meeting of the Association for Computational Linguistics}}.
  \bibinfo{publisher}{Association for Computational Linguistics},
  \bibinfo{address}{Madrid, Spain}, \bibinfo{pages}{96--103}.
\newblock
\urldef\tempurl%
\url{https://doi.org/10.3115/976909.979630}
\showDOI{\tempurl}


\bibitem[\protect\citeauthoryear{Marujo, Gershman, Carbonell, Frederking, and
  Neto}{Marujo et~al\mbox{.}}{2012}]%
        {MARUJO12.672}
\bibfield{author}{\bibinfo{person}{Luís Marujo}, \bibinfo{person}{Anatole
  Gershman}, \bibinfo{person}{Jaime Carbonell}, \bibinfo{person}{Robert
  Frederking}, {and} \bibinfo{person}{JoaÌƒo~P. Neto}.}
  \bibinfo{year}{2012}\natexlab{}.
\newblock \showarticletitle{Supervised Topical Key Phrase Extraction of News
  Stories using Crowdsourcing, Light Filtering and Co-reference Normalization}.
  In \bibinfo{booktitle}{\emph{Proceedings of the Eight International
  Conference on Language Resources and Evaluation (LREC'12)}} (23-25),
  \bibfield{editor}{\bibinfo{person}{Nicoletta Calzolari~(Conference Chair)},
  \bibinfo{person}{Khalid Choukri}, \bibinfo{person}{Thierry Declerck},
  \bibinfo{person}{Mehmet~Uğur Doğan}, \bibinfo{person}{Bente Maegaard},
  \bibinfo{person}{Joseph Mariani}, \bibinfo{person}{Asuncion Moreno},
  \bibinfo{person}{Jan Odijk}, {and} \bibinfo{person}{Stelios Piperidis}}
  (Eds.). \bibinfo{publisher}{European Language Resources Association (ELRA)},
  \bibinfo{address}{Istanbul, Turkey}.
\newblock
\showISBNx{978-2-9517408-7-7}


\bibitem[\protect\citeauthoryear{Meng, Zhao, Han, He, Brusilovsky, and
  Chi}{Meng et~al\mbox{.}}{2017}]%
        {P17-1054}
\bibfield{author}{\bibinfo{person}{Rui Meng}, \bibinfo{person}{Sanqiang Zhao},
  \bibinfo{person}{Shuguang Han}, \bibinfo{person}{Daqing He},
  \bibinfo{person}{Peter Brusilovsky}, {and} \bibinfo{person}{Yu Chi}.}
  \bibinfo{year}{2017}\natexlab{}.
\newblock \showarticletitle{Deep Keyphrase Generation}. In
  \bibinfo{booktitle}{\emph{Proceedings of the 55th Annual Meeting of the
  Association for Computational Linguistics (Volume 1: Long Papers)}}.
  \bibinfo{publisher}{Association for Computational Linguistics},
  \bibinfo{pages}{582--592}.
\newblock
\urldef\tempurl%
\url{https://doi.org/10.18653/v1/P17-1054}
\showDOI{\tempurl}


\bibitem[\protect\citeauthoryear{Mihalcea and Tarau}{Mihalcea and
  Tarau}{2004}]%
        {mihalcea-tarau:2004:EMNLP}
\bibfield{author}{\bibinfo{person}{Rada Mihalcea} {and} \bibinfo{person}{Paul
  Tarau}.} \bibinfo{year}{2004}\natexlab{}.
\newblock \showarticletitle{TextRank: Bringing Order into Texts}. In
  \bibinfo{booktitle}{\emph{Proceedings of EMNLP 2004}},
  \bibfield{editor}{\bibinfo{person}{Dekang Lin} {and} \bibinfo{person}{Dekai
  Wu}} (Eds.). \bibinfo{publisher}{Association for Computational Linguistics},
  \bibinfo{address}{Barcelona, Spain}, \bibinfo{pages}{404--411}.
\newblock


\bibitem[\protect\citeauthoryear{Paszke, Gross, Chintala, Chanan, Yang, DeVito,
  Lin, Desmaison, Antiga, and Lerer}{Paszke et~al\mbox{.}}{2017}]%
        {paszke2017automatic}
\bibfield{author}{\bibinfo{person}{Adam Paszke}, \bibinfo{person}{Sam Gross},
  \bibinfo{person}{Soumith Chintala}, \bibinfo{person}{Gregory Chanan},
  \bibinfo{person}{Edward Yang}, \bibinfo{person}{Zachary DeVito},
  \bibinfo{person}{Zeming Lin}, \bibinfo{person}{Alban Desmaison},
  \bibinfo{person}{Luca Antiga}, {and} \bibinfo{person}{Adam Lerer}.}
  \bibinfo{year}{2017}\natexlab{}.
\newblock \showarticletitle{Automatic differentiation in pytorch}. In
  \bibinfo{booktitle}{\emph{NIPS 2017 Workshop Autodiff}}.
\newblock


\bibitem[\protect\citeauthoryear{Salton and Buckley}{Salton and
  Buckley}{1988}]%
        {SALTON1988513}
\bibfield{author}{\bibinfo{person}{Gerard Salton} {and}
  \bibinfo{person}{Christopher Buckley}.} \bibinfo{year}{1988}\natexlab{}.
\newblock \showarticletitle{Term-weighting approaches in automatic text
  retrieval}.
\newblock \bibinfo{journal}{\emph{Information Processing \& Management}}
  \bibinfo{volume}{24}, \bibinfo{number}{5} (\bibinfo{year}{1988}),
  \bibinfo{pages}{513 -- 523}.
\newblock
\showISSN{0306-4573}
\urldef\tempurl%
\url{https://doi.org/10.1016/0306-4573(88)90021-0}
\showDOI{\tempurl}


\bibitem[\protect\citeauthoryear{Schutz}{Schutz}{2008}]%
        {schutz2008keyphrase}
\bibfield{author}{\bibinfo{person}{Alexander~Thorsten Schutz}.}
  \bibinfo{year}{2008}\natexlab{}.
\newblock \showarticletitle{Keyphrase extraction from single documents in the
  open domain exploiting linguistic and statistical methods}.
\newblock \bibinfo{journal}{\emph{Master's thesis, National University of
  Ireland}} (\bibinfo{year}{2008}).
\newblock


\bibitem[\protect\citeauthoryear{Sood, Owsley, Hammond, and Birnbaum}{Sood
  et~al\mbox{.}}{2007}]%
        {DBLP:conf/icwsm/SoodOHB07}
\bibfield{author}{\bibinfo{person}{Sanjay Sood}, \bibinfo{person}{Sara Owsley},
  \bibinfo{person}{Kristian~J. Hammond}, {and} \bibinfo{person}{Larry
  Birnbaum}.} \bibinfo{year}{2007}\natexlab{}.
\newblock \showarticletitle{TagAssist: Automatic Tag Suggestion for Blog
  Posts}. In \bibinfo{booktitle}{\emph{Proceedings of the First International
  Conference on Weblogs and Social Media, {ICWSM} 2007, Boulder, Colorado, USA,
  March 26-28, 2007}}.
\newblock
\urldef\tempurl%
\url{http://www.icwsm.org/papers/paper10.html}
\showURL{%
\tempurl}


\bibitem[\protect\citeauthoryear{Sutskever, Vinyals, and Le}{Sutskever
  et~al\mbox{.}}{2014}]%
        {NIPS2014_5346}
\bibfield{author}{\bibinfo{person}{Ilya Sutskever}, \bibinfo{person}{Oriol
  Vinyals}, {and} \bibinfo{person}{Quoc~V Le}.}
  \bibinfo{year}{2014}\natexlab{}.
\newblock \showarticletitle{Sequence to Sequence Learning with Neural
  Networks}.
\newblock In \bibinfo{booktitle}{\emph{Advances in Neural Information
  Processing Systems 27}}, \bibfield{editor}{\bibinfo{person}{Z.~Ghahramani},
  \bibinfo{person}{M.~Welling}, \bibinfo{person}{C.~Cortes},
  \bibinfo{person}{N.~D. Lawrence}, {and} \bibinfo{person}{K.~Q. Weinberger}}
  (Eds.). \bibinfo{publisher}{Curran Associates, Inc.},
  \bibinfo{pages}{3104--3112}.
\newblock
\urldef\tempurl%
\url{http://papers.nips.cc/paper/5346-sequence-to-sequence-learning-with-neural-networks.pdf}
\showURL{%
\tempurl}


\bibitem[\protect\citeauthoryear{Teneva and Cheng}{Teneva and Cheng}{2017}]%
        {teneva-cheng:2017:Short}
\bibfield{author}{\bibinfo{person}{Nedelina Teneva} {and}
  \bibinfo{person}{Weiwei Cheng}.} \bibinfo{year}{2017}\natexlab{}.
\newblock \showarticletitle{Salience Rank: Efficient Keyphrase Extraction with
  Topic Modeling}. In \bibinfo{booktitle}{\emph{Proceedings of the 55th Annual
  Meeting of the Association for Computational Linguistics (Volume 2: Short
  Papers)}}. \bibinfo{publisher}{Association for Computational Linguistics},
  \bibinfo{address}{Vancouver, Canada}, \bibinfo{pages}{530--535}.
\newblock
\urldef\tempurl%
\url{http://aclweb.org/anthology/P17-2084}
\showURL{%
\tempurl}


\bibitem[\protect\citeauthoryear{Tu, Lu, Liu, Liu, and Li}{Tu
  et~al\mbox{.}}{2016}]%
        {tu-etal-2016-modeling}
\bibfield{author}{\bibinfo{person}{Zhaopeng Tu}, \bibinfo{person}{Zhengdong
  Lu}, \bibinfo{person}{Yang Liu}, \bibinfo{person}{Xiaohua Liu}, {and}
  \bibinfo{person}{Hang Li}.} \bibinfo{year}{2016}\natexlab{}.
\newblock \showarticletitle{Modeling Coverage for Neural Machine Translation}.
  In \bibinfo{booktitle}{\emph{Proceedings of the 54th Annual Meeting of the
  Association for Computational Linguistics (Volume 1: Long Papers)}}.
  \bibinfo{publisher}{Association for Computational Linguistics},
  \bibinfo{address}{Berlin, Germany}, \bibinfo{pages}{76--85}.
\newblock
\urldef\tempurl%
\url{https://doi.org/10.18653/v1/P16-1008}
\showDOI{\tempurl}


\bibitem[\protect\citeauthoryear{Wan and Xiao}{Wan and Xiao}{2008}]%
        {Wan:2008:SDK:1620163.1620205}
\bibfield{author}{\bibinfo{person}{Xiaojun Wan} {and} \bibinfo{person}{Jianguo
  Xiao}.} \bibinfo{year}{2008}\natexlab{}.
\newblock \showarticletitle{Single Document Keyphrase Extraction Using
  Neighborhood Knowledge}. In \bibinfo{booktitle}{\emph{Proceedings of the 23rd
  National Conference on Artificial Intelligence - Volume 2}}
  \emph{(\bibinfo{series}{AAAI'08})}. \bibinfo{publisher}{AAAI Press},
  \bibinfo{pages}{855--860}.
\newblock
\showISBNx{978-1-57735-368-3}
\urldef\tempurl%
\url{http://dl.acm.org/citation.cfm?id=1620163.1620205}
\showURL{%
\tempurl}


\bibitem[\protect\citeauthoryear{Wang, Liu, and McDonald}{Wang
  et~al\mbox{.}}{2014}]%
        {10.1007/978-3-642-54906-9_14}
\bibfield{author}{\bibinfo{person}{Rui Wang}, \bibinfo{person}{Wei Liu}, {and}
  \bibinfo{person}{Chris McDonald}.} \bibinfo{year}{2014}\natexlab{}.
\newblock \showarticletitle{How Preprocessing Affects Unsupervised Keyphrase
  Extraction}. In \bibinfo{booktitle}{\emph{Computational Linguistics and
  Intelligent Text Processing}}, \bibfield{editor}{\bibinfo{person}{Alexander
  Gelbukh}} (Ed.). \bibinfo{publisher}{Springer Berlin Heidelberg},
  \bibinfo{address}{Berlin, Heidelberg}, \bibinfo{pages}{163--176}.
\newblock
\showISBNx{978-3-642-54906-9}


\bibitem[\protect\citeauthoryear{Wang, Liu, and McDonald}{Wang
  et~al\mbox{.}}{2015}]%
        {10.1007/978-3-319-19548-3_21}
\bibfield{author}{\bibinfo{person}{Rui Wang}, \bibinfo{person}{Wei Liu}, {and}
  \bibinfo{person}{Chris McDonald}.} \bibinfo{year}{2015}\natexlab{}.
\newblock \showarticletitle{Using Word Embeddings to Enhance Keyword
  Identification for Scientific Publications}. In
  \bibinfo{booktitle}{\emph{Databases Theory and Applications}},
  \bibfield{editor}{\bibinfo{person}{Mohamed~A. Sharaf},
  \bibinfo{person}{Muhammad~Aamir Cheema}, {and} \bibinfo{person}{Jianzhong
  Qi}} (Eds.). \bibinfo{publisher}{Springer International Publishing},
  \bibinfo{address}{Cham}, \bibinfo{pages}{257--268}.
\newblock
\showISBNx{978-3-319-19548-3}


\bibitem[\protect\citeauthoryear{Witten, Bainbridge, and Nichols}{Witten
  et~al\mbox{.}}{2009}]%
        {witten2009build}
\bibfield{author}{\bibinfo{person}{Ian~H Witten}, \bibinfo{person}{David
  Bainbridge}, {and} \bibinfo{person}{David~M Nichols}.}
  \bibinfo{year}{2009}\natexlab{}.
\newblock \bibinfo{booktitle}{\emph{How to build a digital library}}.
\newblock \bibinfo{publisher}{Morgan Kaufmann}.
\newblock


\bibitem[\protect\citeauthoryear{Witten, Paynter, Frank, Gutwin, and
  Nevill-Manning}{Witten et~al\mbox{.}}{1999}]%
        {Witten:1999:KPA:313238.313437}
\bibfield{author}{\bibinfo{person}{Ian~H. Witten}, \bibinfo{person}{Gordon~W.
  Paynter}, \bibinfo{person}{Eibe Frank}, \bibinfo{person}{Carl Gutwin}, {and}
  \bibinfo{person}{Craig~G. Nevill-Manning}.} \bibinfo{year}{1999}\natexlab{}.
\newblock \showarticletitle{KEA: Practical Automatic Keyphrase Extraction}. In
  \bibinfo{booktitle}{\emph{Proceedings of the Fourth ACM Conference on Digital
  Libraries}} \emph{(\bibinfo{series}{DL '99})}. \bibinfo{publisher}{ACM},
  \bibinfo{address}{New York, NY, USA}, \bibinfo{pages}{254--255}.
\newblock
\showISBNx{1-58113-145-3}
\urldef\tempurl%
\url{https://doi.org/10.1145/313238.313437}
\showDOI{\tempurl}


\bibitem[\protect\citeauthoryear{Ye and Wang}{Ye and Wang}{2018}]%
        {ye-wang:2018:EMNLP}
\bibfield{author}{\bibinfo{person}{Hai Ye} {and} \bibinfo{person}{Lu Wang}.}
  \bibinfo{year}{2018}\natexlab{}.
\newblock \showarticletitle{Semi-Supervised Learning for Neural Keyphrase
  Generation}. In \bibinfo{booktitle}{\emph{Proceedings of the 2018 Conference
  on Empirical Methods in Natural Language Processing}}.
  \bibinfo{publisher}{Association for Computational Linguistics},
  \bibinfo{address}{Brussels, Belgium}, \bibinfo{pages}{4142--4153}.
\newblock
\urldef\tempurl%
\url{http://www.aclweb.org/anthology/D18-1447}
\showURL{%
\tempurl}


\bibitem[\protect\citeauthoryear{Zesch and Gurevych}{Zesch and
  Gurevych}{2009}]%
        {zesch-gurevych:2009:RANLP09}
\bibfield{author}{\bibinfo{person}{Torsten Zesch} {and} \bibinfo{person}{Iryna
  Gurevych}.} \bibinfo{year}{2009}\natexlab{}.
\newblock \showarticletitle{Approximate Matching for Evaluating Keyphrase
  Extraction}. In \bibinfo{booktitle}{\emph{Proceedings of the International
  Conference RANLP-2009}}. \bibinfo{publisher}{Association for Computational
  Linguistics}, \bibinfo{address}{Borovets, Bulgaria},
  \bibinfo{pages}{484--489}.
\newblock
\urldef\tempurl%
\url{http://www.aclweb.org/anthology/R09-1086}
\showURL{%
\tempurl}


\bibitem[\protect\citeauthoryear{Zhai}{Zhai}{1997}]%
        {zhai-1997-fast}
\bibfield{author}{\bibinfo{person}{Chengxiang Zhai}.}
  \bibinfo{year}{1997}\natexlab{}.
\newblock \showarticletitle{Fast Statistical Parsing of Noun Phrases for
  Document Indexing}. In \bibinfo{booktitle}{\emph{Fifth Conference on Applied
  Natural Language Processing}}. \bibinfo{publisher}{Association for
  Computational Linguistics}, \bibinfo{address}{Washington, DC, USA},
  \bibinfo{pages}{312--319}.
\newblock
\urldef\tempurl%
\url{https://doi.org/10.3115/974557.974603}
\showDOI{\tempurl}


\end{thebibliography}

\end{document}